\documentclass[final,1p,times]{elsarticle}

\usepackage{amssymb}

\journal{Nuclear Physics A}

\begin{document}

\begin{frontmatter}



\title{Kaonic hydrogen X-ray measurement in SIDDHARTA}


\author[lnf]{M.~Bazzi}
\author[victoria]{G.~Beer}
\author[milano]{L.~Bombelli}
\author[lnf,ifin]{A.M.~Bragadireanu}
\author[smi]{M.~Cargnelli}
\author[lnf]{G.~Corradi}
\author[lnf]{C.~Curceanu (Petrascu)}
\author[lnf]{A.~d'Uffizi}
\author[milano]{C.~Fiorini}
\author[milano]{T.~Frizzi}
\author[roma]{F.~Ghio}
\author[lnf]{C.~Guaraldo}
\author[ut]{R.S.~Hayano}
\author[lnf,ifin]{M.~Iliescu}
\author[smi]{T.~Ishiwatari}
\author[riken]{M.~Iwasaki}
\author[smi,tum]{P.~Kienle}
\author[lnf]{P.~Levi~Sandri}
\author[milano]{A.~Longoni}
\author[lnf]{V.~Lucherini}
\author[smi]{J.~Marton}
\author[lnf]{S.~Okada\corref{cor}}
\ead{shinji.okada@lnf.infn.it}
\author[lnf,ifin]{D.~Pietreanu}
\author[ifin]{T.~Ponta}
\author[lnf]{A.~Rizzo}
\author[lnf]{A.~Romero~Vidal}
\author[lnf]{A.~Scordo}
\author[ut]{H.~Shi}
\author[lnf,ifin]{D.L.~Sirghi}
\author[lnf,ifin]{F.~Sirghi}
\author[ut]{H.~Tatsuno\fnref{nowlnf}}
\author[ifin]{A.~Tudorache}
\author[ifin]{V.~Tudorache}
\author[lnf]{O.~Vazquez~Doce}
\author[smi]{E.~Widmann}
\author[smi]{J.~Zmeskal}
\cortext[cor]{Corresponding author.}
 
\address[lnf]{INFN, Laboratori Nazionali di Frascati, Frascati (Roma), Italy}
\address[victoria]{Department of Physics and Astronomy,
 University of Victoria, Victoria, BC, Canada}
\address[milano]{Politecnico di Milano, Sezione di Elettronica, Milano, Italy}
\address[ifin]{Horia Hulubei National Institute of Physics
 and Nuclear Engineering, Magurele, Romania}
\address[smi]{Stefan-Meyer-Institut f$\ddot{\mbox{u}}$r
 subatomare Physik, Vienna, Austria}
\address[roma]{INFN Sezione di Roma I and Istituto Superiore di
 Sanit$\grave{\mbox{a}}$, Roma, Italy}
\address[ut]{University of Tokyo, Tokyo, Japan}
\address[riken]{RIKEN, The Institute of Physics and Chemical Research,
 Saitama, Japan}
\address[tum]{Technische Universit$\ddot{\mbox{a}}$t
 M$\ddot{\mbox{u}}$nchen, Physik Department, Garching, Germany}

\fntext[nowlnf]{Present address: 
 INFN, Laboratori Nazionali di Frascati, Frascati (Roma), Italy}

\begin{abstract}
Kaonic hydrogen atoms provide a unique laboratory to probe
the kaon-nucleon strong interaction at the energy threshold,
allowing an investigation of the interplay between spontaneous
and explicit chiral symmetry breaking in low-energy QCD. 
The SIDDHARTA Collaboration has measured the $K$-series X rays of
kaonic hydrogen atoms at the DA$\Phi$NE electron-positron collider of
Laboratori Nazionali di Frascati,
and has determined
the most precise values of the strong-interaction induced shift and width 
of the $1s$ atomic energy level.
This result provides
vital constraints on the theoretical
description of the low-energy $\overline{K}N$ interaction.
\end{abstract}


\end{frontmatter}


\section{Introduction}
\label{sec:intro}

Kaonic hydrogen is an exotic atom composed of a proton and a
$K^-$ bound by the Coulomb force.
The strong interaction shifts the $1s$ atomic energy level
from its pure electromagnetic (EM) value
and broadens it as a result of the shortened lifetime due to absorption.
Since the strong-interaction effects for the higher states
 ($i.e.$, $2p$, $3p$ ...) are negligible in comparison,
the $1s$ shift and width
can be deduced from the spectroscopy of kaonic-hydrogen X-ray transitions
feeding the $1s$ states, namely the $K$-series X rays.

\begin{figure}[t]
 \begin{center}
  \includegraphics*[width=1.0\linewidth]{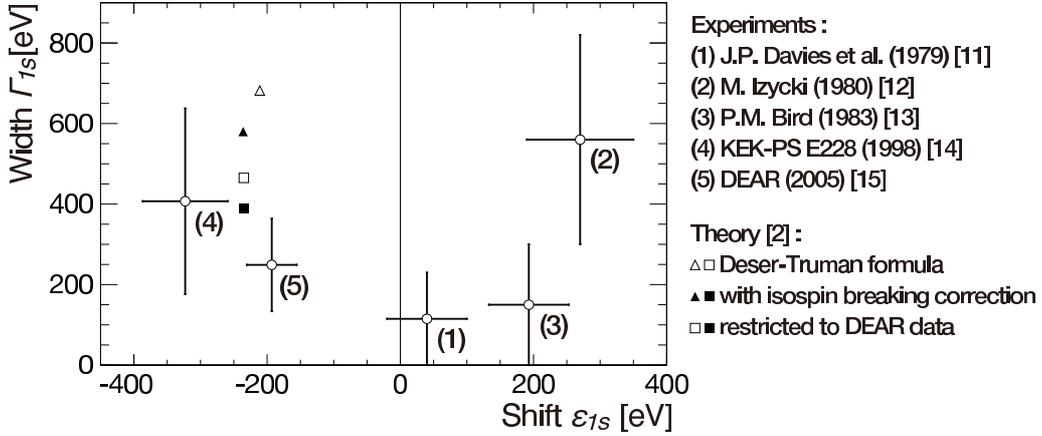}
  \caption{Comparison of previous experimental and theoretical results for the
  strong-interaction $1s$-energy-level shift
  and width of kaonic hydrogen.
  The open circles with error bars are experimental results:
  J.~D.~Davies {\em et al.} (1979) \cite{J_D_Davies_1979},
  M.~Izycki {\em et al.} (1980) \cite{M_Izycki_1980},
  P.~M.~Bird, {\em et al.} (1983) \cite{P_M_Bird_1983},
  KEK-PS E228 (1997) \cite{kpx} and DEAR (2005) \cite{dear}.
  The error bars correspond to quadratically added statistical and
  systematic errors.
  The symbols without error bars show the values
  of a theoretical calculation \cite{Borasoy2005}
  both by using the Deser-Trueman formula (empty symbols)
  and by including isospin breaking corrections (filled symbols).
  The fit restricted to the DEAR data is represented
  by the squares.
  }
  \label{fig:Comparison}
 \end{center}
\end{figure}

The measured strong-interaction shift $\epsilon_{1s}$ and width
$\Gamma_{1s}$ are directly related to
the real and imaginary parts of the complex $K^-p$ $S$-wave scattering
length $a_{K^-p}$ which, in the isospin limit, is given
by the Deser-Truemann formula \cite{Deser_Trueman} :
\begin{displaymath}
 \epsilon_{1s} + \frac{i}{2} \Gamma_{1s} = 2 \alpha^3 \mu^2 a_{K^-p}
  = 412 ~\mbox{eV fm}^{-1}~ a_{K^-p}
\end{displaymath} 
where $\mu$ is the reduced mass of the $K^-p$ system and
$\alpha$ is the fine structure constant.
The kaonic-hydrogen X-ray data are therefore crucial for theories of
the $\overline{K}N$ system together with the low-energy
$\overline{K}N$ data.
Note that recent Coulomb and isospin breaking corrections to the formula
turn out to be important, as first shown in \cite{Borasoy2005}.
Fig.~\ref{fig:Comparison}
shows the comparison between uncorrected and corrected values in this
theoretical calculation \cite{Borasoy2005}.

The low-energy $\overline{K}N$ system has attracted attention as a
sensitive testing ground for chiral SU(3) dynamics in low-energy QCD,
allowing investigation of the interplay between spontaneous
and explicit chiral symmetry breaking due to the relatively large
strange quark mass
which plays an intermediate role between ``light'' and ``heavy''.
The data are also strongly related to recent hot topics
-- the structure of the $\Lambda(1405)$ resonance 
($e.g.,$ \cite{twopole,hyodo,hj})
and the deeply bound kaonic systems ($e.g.,$ \cite{AY,FINUDA,KEK,DISTO}).
Recent progress in this field is summarized in \cite{ECT}.

Historically there were several measurements 
of the strong-interaction shift $\epsilon_{1s}$
\footnote{Note that $\epsilon_{1s}$ is defined as $\epsilon_{1s} \equiv - (E_{1s}
- E_{1s}^{EM})$, where $E_{1s}$ is the energy of the $1s$ level and
$E_{1s}^{EM}$ is the energy calculated using only the EM interaction.}
and width $\Gamma_{1s}$
\cite{J_D_Davies_1979, M_Izycki_1980, P_M_Bird_1983, kpx, dear}.
In the 1970s and the 1980s
three groups \cite{J_D_Davies_1979, M_Izycki_1980, P_M_Bird_1983}
reported a measured attractive shift
(positive $\epsilon_{1s}$) as shown in Fig.~\ref{fig:Comparison},
while the information extracted from the analyses of the
low energy $\overline{K}N$ data 
($e.g.,$ \cite{Humphrey1962, Kim1967, Martin1981}) shows
a repulsive shift (negative $\epsilon_{1s}$).
This contradiction has been known as the ``kaonic hydrogen puzzle''.

In 1997, the first distinct peaks of the kaonic-hydrogen X rays
were observed by the KEK-PS E228 group \cite{kpx}
with a significant improvement in the signal-to-background ratio
by the use of a gaseous hydrogen target,
where previous experiments had employed liquid hydrogen.
It was crucial to use a low-density target,
namely a gaseous target,
because the X-ray yields quickly decrease
towards higher density due to the Stark mixing effect.
The observed repulsive shift
was consistent in sign with the analysis of
the low energy $\overline{K}N$ scattering data,
resolving the long-standing discrepancy.

More recent values reported by the DEAR group in 2005 \cite{dear},
with substantially reduced errors as shown in Fig.~\ref{fig:Comparison},
firmly established the repulsive shift obtained in the
previous E228 experiment.

Intensive theoretical studies were performed based on these DEAR results
\cite{Borasoy2005,correct,nissler,oller,oller2,bor,oller3,revai,cieply,shevchenko,oset}
\footnote{See also \cite{NPA2010,PPNP}.}.
However, most calculations show difficulties in explaining
all the experimental results in a consistent way.
As an example,
the theoretical values with and without restriction of DEAR data
are plotted in Fig.~\ref{fig:Comparison} \cite{Borasoy2005}.

In the present SIDDHARTA experiment we have determined the most precise
values to date of the $1s$ strong interaction shift and width of kaonic
hydrogen \cite{SIDT_KH} in an attempt to clarify this difficulty.

\section{Experiment}
\label{sec:exp}

The SIDDHARTA experiment was performed
at the recently upgraded DA$\Phi$NE
positron-electron collider \cite{Milardi,Zobov}.
The collider produces $\phi$-resonances, 49 \% of which decay into
back-to-back $K^+ K^-$ pairs.
The resulting monochromatic low-energy kaons are efficiently
stopped in a cryogenic hydrogen ``gaseous'' target.

\begin{figure}[t]
  \begin{center}
   \includegraphics*[width=0.5\linewidth]{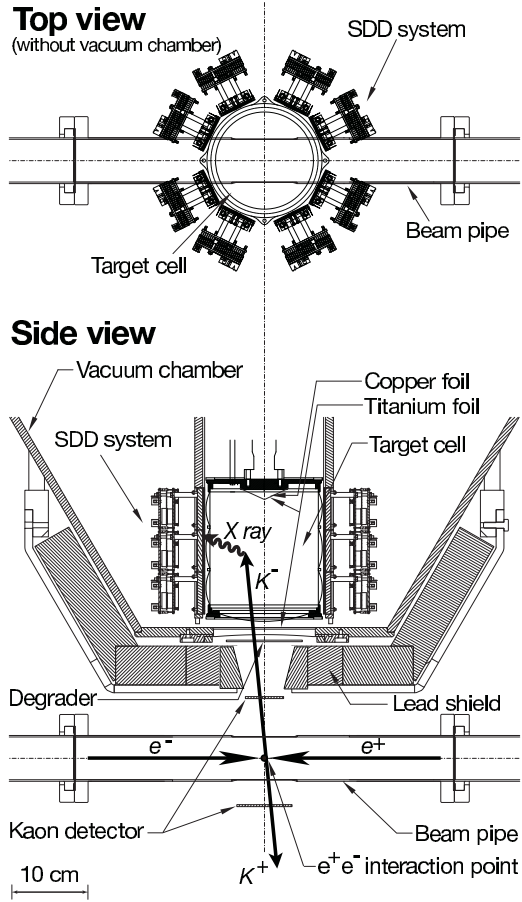}
   \caption{A schematic view of the SIDDHARTA setup
   installed at the e$^+$e$^-$ interaction point of DA$\rm{\Phi}$NE.}
   \label{fig:Setup}
  \end{center}
\end{figure}

A schematic view of the SIDDHARTA setup is shown in
Fig.~\ref{fig:Setup}.
A coincidence of two plastic scintillation counters mounted above and
below the $e^+ e^-$ interaction point was used as a kaon trigger.
X rays emitted from the kaonic atoms were detected by 144 silicon
drift detectors (SDDs), each having an effective area of 1 cm$^2$
and a thickness of 450 $\mu$m,
developed within a European research project devoted to this
experiment.
The SDDs had an energy
resolution of $\sim$~180 eV (FWHM) at 8 keV
and timing resolution below 1 $\mu$sec,
in contrast to the CCD detectors used in DEAR \cite{DEAR-CCD}
which had no timing capability.
In comparison with DEAR,
the main source of background coming from beam losses was highly suppressed.
A detailed description of our experimental setup is given in 
\cite{SIDT_KH,SIDT_KHe3,SIDT_KHe4}.

\section{Data analysis}
\label{sec:ana}

\begin{figure}[t]
 \begin{center}
  \includegraphics*[width=0.7\columnwidth]{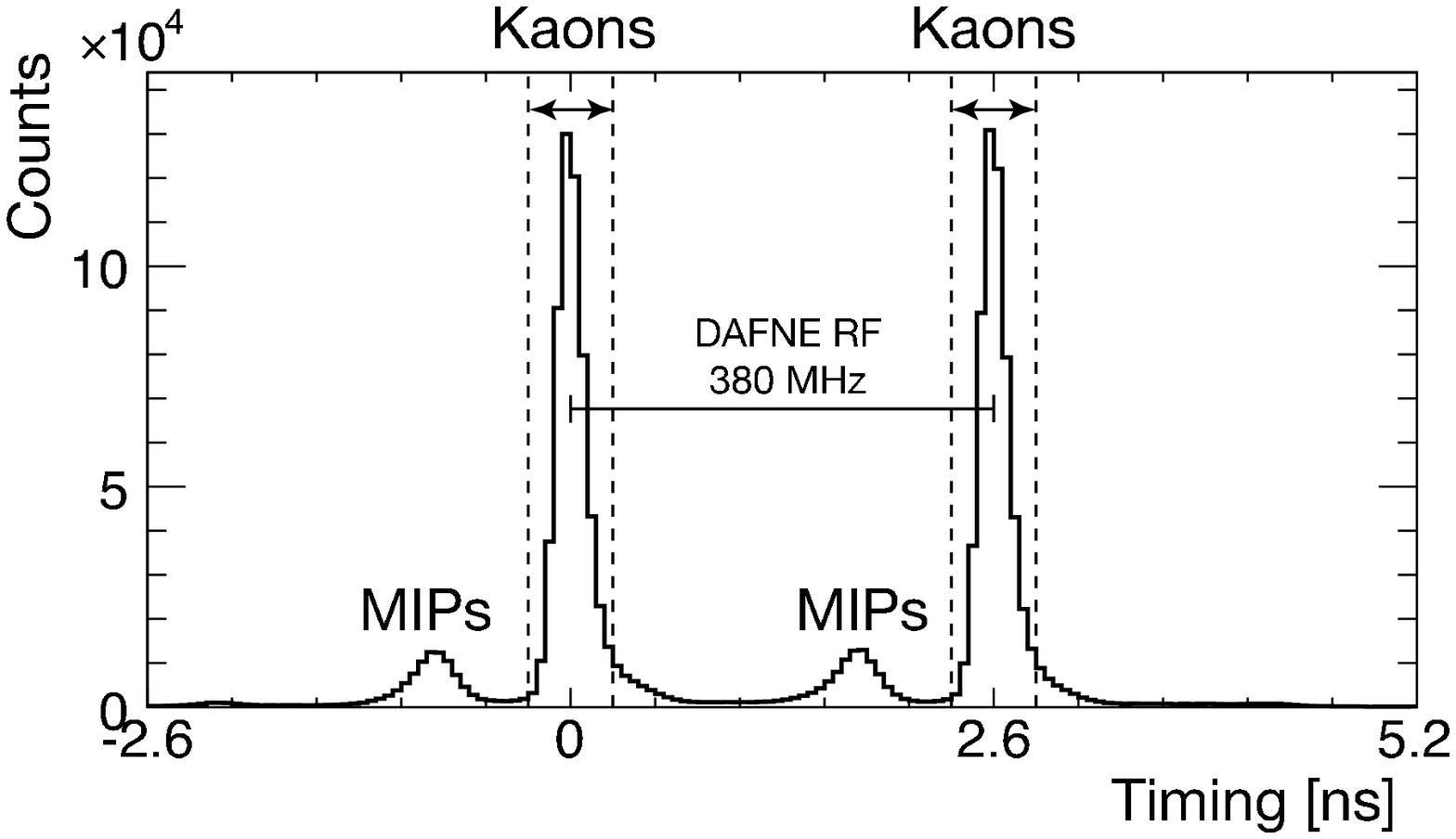}
  \caption{Kaon identification
  using timing of the coincidence signals in the
  kaon detector with respect to the RF signal of $\sim$ 368.7 MHz
  from DA$\Phi$NE.
  }
  \label{fig:KID}
 \end{center}
\end{figure}

Fig.~\ref{fig:KID} shows
the timing distribution of the coincidence signals in the kaon detector
with respect to the $\sim$~368.7 MHz RF signal from DA$\Phi$NE.
The spectrum shows clearly that kaon events can be separated from 
minimum ionizing particles
by setting a time gate as indicated by arrows in the figure.
A half frequency of the beam synchronous timing signal was used
for the start RF timing --
therefore there are two identical coincidence signals
in the spectrum at an interval of $\sim$ 2.6 ns.

\begin{figure}[t]
 \begin{center}
  \includegraphics*[width=0.7\columnwidth]{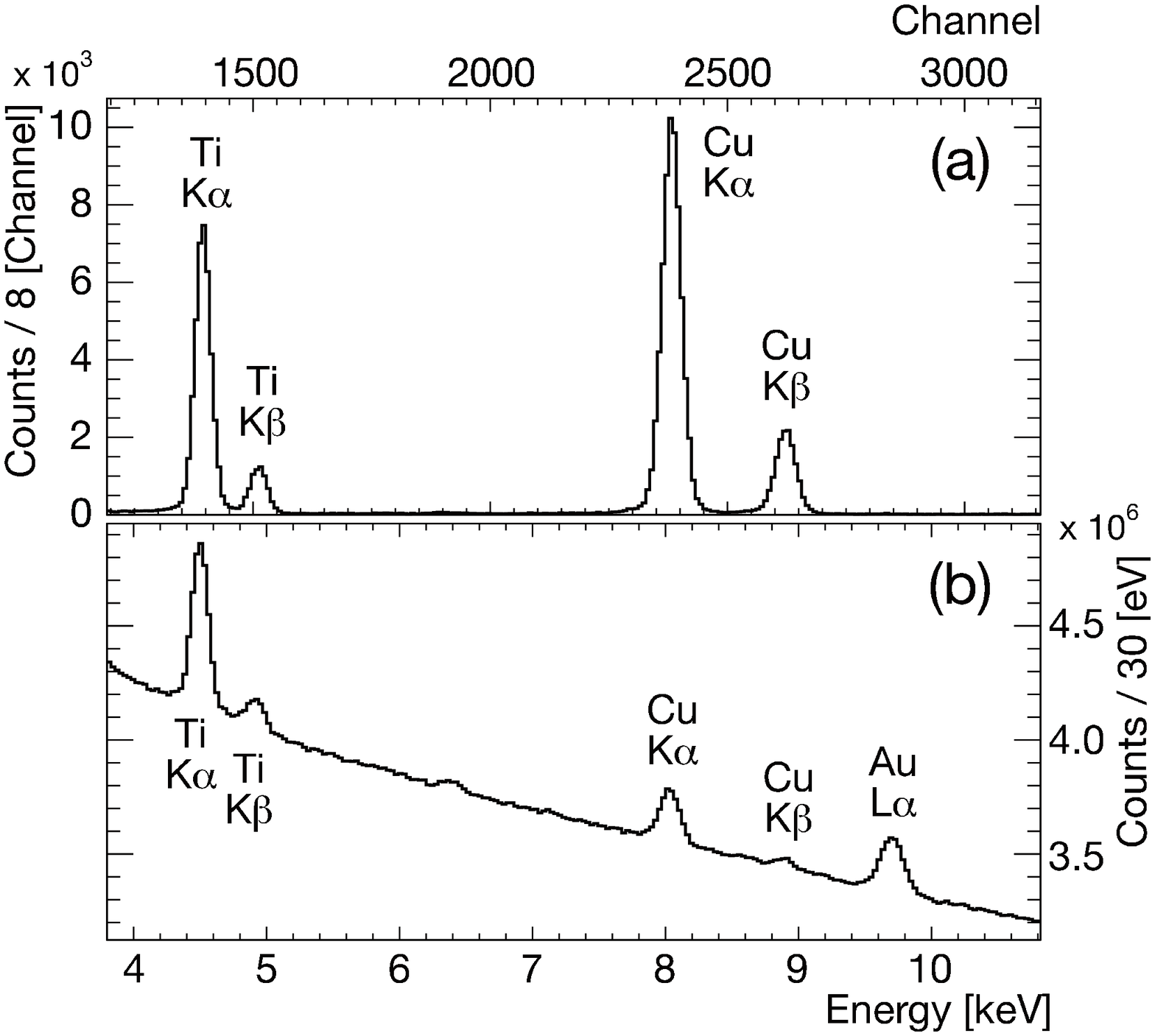}
  \caption{X-ray energy spectra for
  (a) data taken with the X-ray tube, and
  (b) data uncorrelated to the kaon production timing.
  The refined {\it in-situ} calibration 
  in gain (energy) and resolution (response shape) 
  of the summed spectrum of all SDDs was obtained using titanium,
  copper, and gold characteristic X-ray lines in the spectrum (b).
  }
  \label{fig:CalibSelfSpec}
 \end{center}
\end{figure}

Energy calibration of each SDD's ADC spectrum
was performed by periodic measurements
of characteristic X-ray lines from titanium and copper foils
located as shown in the setup (Fig.~\ref{fig:Setup})
excited by an X-ray tube.
Fig.~\ref{fig:CalibSelfSpec} (a) shows 
a typical X-ray spectrum from the X-ray tube data for a single SDD.
Characteristic X rays of titanium and copper were obtained
with high statistics.
A remote-controlled system moved the kaon detector out and the X-ray
tube in for these calibration measurements, once every $\sim$ 4 hours.
The energy scale was calibrated linearly by the two $K\alpha$ lines.

The refined {\it in-situ} calibration in energy was performed
using characteristic X-ray lines of
titanium, copper and gold ($L\alpha$)
excited by the uncorrelated background without trigger
in the summed spectrum of all SDDs.
The spectrum for the complete statistics
of the kaonic-hydrogen dataset
is shown in Fig.~\ref{fig:CalibSelfSpec} (b).
Moreover the parameters of the (energy-dependent) energy resolution
for the summed spectrum were also evaluated
using those peaks in this {\it in-situ} energy spectrum
along with the kaonic carbon lines from wall stops
in the final energy spectrum in the kaon-triggered mode.

\begin{figure}[htpb]
\centering
\includegraphics[width=0.7\columnwidth,angle=0]{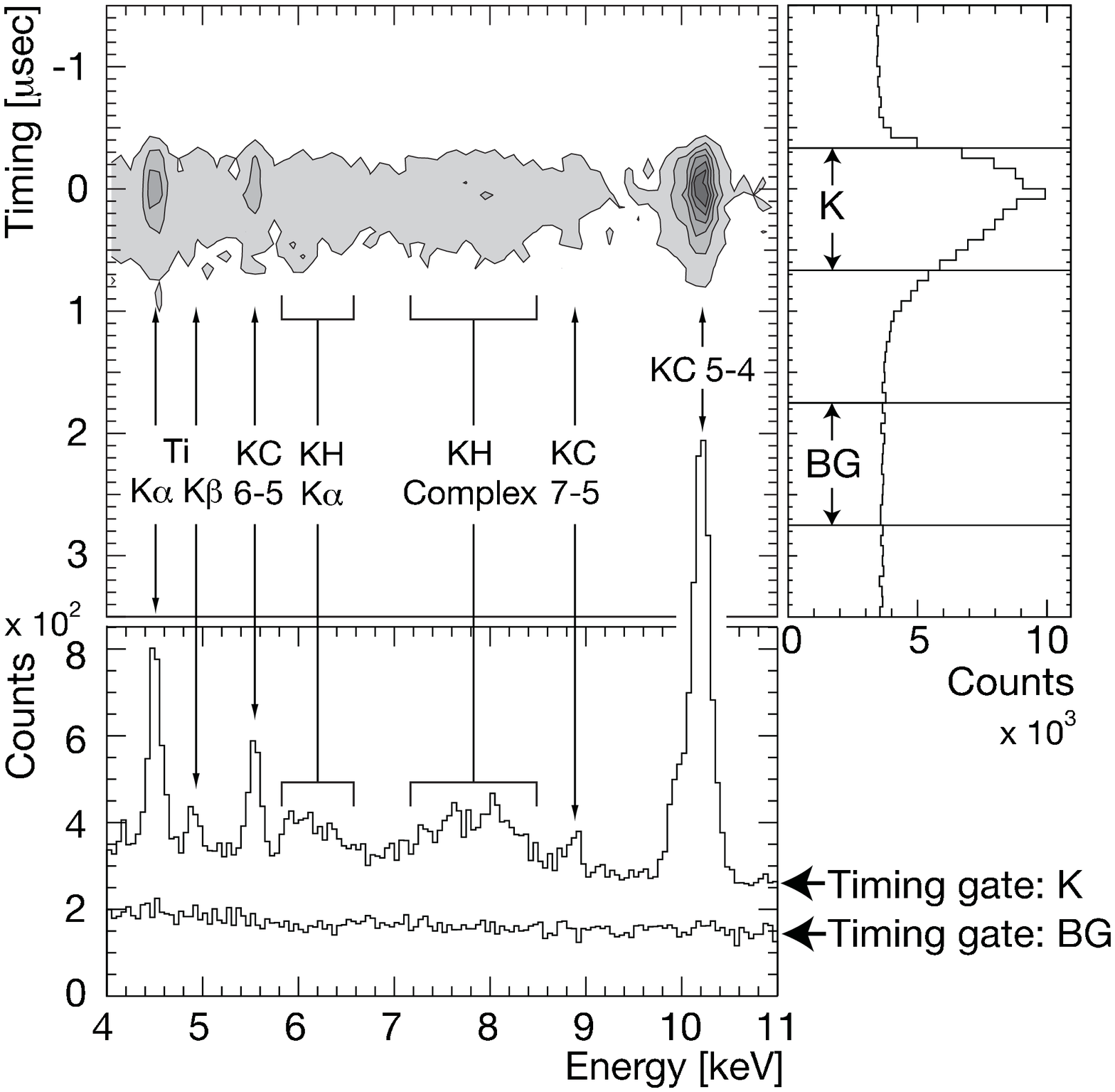} 
 \caption{Measured X-ray time and energy spectra for $K^-$ triggered
 events of hydrogen data.
 The top-left panel shows a correlation plot of the X-ray energy
 measured by SDDs vs the time difference between kaon arrival and
 X-ray detection,
 where a time-walk correction was applied.
 The projections on the time and energy axes are also shown  
 at the right and bottom.
 By selecting kaon and background time gates
 indicated with ``K'' and ``BG'' in the timing spectrum,
 both the energy spectra are displayed.
 }
 \label{fig:EneTimeSpec}
\end{figure}

A correlation plot of the X-ray energy measured by SDDs
vs the time difference between kaon arrival (with kaon detectors)
and X-ray detection (with SDDs)
for hydrogen data is shown in Fig.~\ref{fig:EneTimeSpec}.
The horizontal band is due to the kaon-induced events.
Its projected time and energy spectra are also displayed in the figure.
A typical width of the time-correlation, after the time-walk correction,
was about 800 ns (FWHM) which reflected the drift-time distribution
of the electrons in the SDD.
The kaon and background time gates, each having a width of 1 $\mu$sec,
are indicated therein with arrows.

The continuous background is related to the following two types of
particles: the charged kaon secondaries (prompt background)
and lost beam particles (accidental background).
Comparing the energy spectra of data with SDD timing gates set to
``K'' (kaon) and ``BG'' (background) in Fig.~\ref{fig:EneTimeSpec},
the prompt background is at the same level as
the accidental background.
In the most recent previous measurement
of kaonic hydrogen X rays (DEAR) \cite{dear},
also performed at DA$\Phi$NE,
the kaonic-hydrogen spectrum suffered from the huge accidental
background due to lack of the timing capability of the X-ray detectors
(CCDs) used.
The event selection using the time information
significantly reduced the accidental background and improved the
signal-to-background ratio by more than a factor of 10 with respect to
the corresponding DEAR ratio of about 1/100.

\begin{figure}[t]
 \begin{center}
  \includegraphics*[width=1.0\columnwidth]{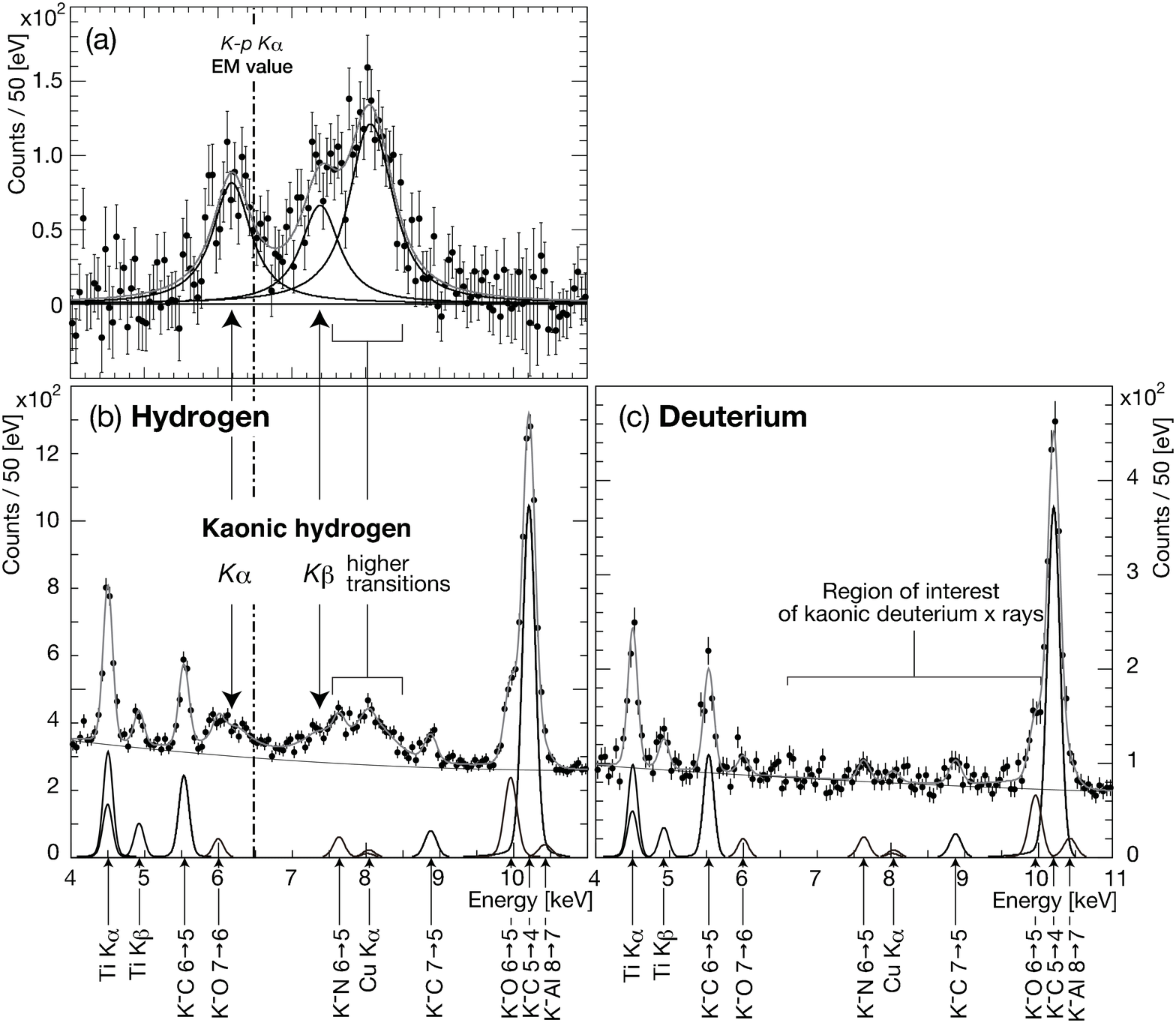}
  \caption{A global simultaneous fit result of
  the X-ray energy spectra of hydrogen and deuterium data.
  (a) Residuals of the measured kaonic-hydrogen X-ray spectrum
  after subtraction of the fitted background,
  clearly displaying the kaonic-hydrogen $K$-series transitions.
  The fit components of the $K^-p$ transitions are also shown,
  where the higher transitions, greater than $K\beta$, are summed.
  (b)(c) Measured energy spectra with the fit lines for each dataset.
  Fit components of the background X-ray lines
  and a continuous background are also shown.
  The dot-dashed vertical line indicates
  the EM value of the kaonic-hydrogen $K\alpha$ energy.
  (Note that the characteristic $K\alpha$ line consists of
  $K\alpha1$ and $K\alpha2$ lines, both of which are shown.)
  }
 \label{fig:SpecFit}
 \end{center}
\end{figure}

Fig.~\ref{fig:SpecFit} (b) shows the kaonic-hydrogen X-ray spectrum.
We have also measured an X-ray spectrum with a deuterium target (for
the first-ever exploratory measurement of kaonic-deuterium X rays),
as shown in Fig.~\ref{fig:SpecFit} (c).
The kaonic-hydrogen X-ray transitions were clearly observed while
those for kaonic deuterium were not visible.
This appears to be consistent with the theoretical expectation 
that kaonic deuterium X rays have one order lower yield per stopped
$K^-$ and greater width than those of kaonic hydrogen X rays
($e.g.,$ \cite{T_Koike_1996}).

A dot-dashed line in Fig.~\ref{fig:SpecFit} (b) indicates the EM value of the
kaonic-hydrogen $K\alpha$.
Comparing the kaonic-hydrogen $K\alpha$ peak and the EM value, there
is no room for doubt about a repulsive shift of the
kaonic-hydrogen $1s$-energy level,
which is consistent with the analysis of the low energy
$\overline{K}N$ scattering data.
 
Many other kaonic-atom X rays and characteristic X rays were detected
in both spectra as indicated with arrows in the figures.
Those kaonic-atom lines are attributable to the target-cell wall made
of Kapton polyimide film (C$_{22}$H$_{10}$O$_5$N$_2$) and its support
frames made of aluminum.
The characteristic X rays come from high-purity titanium and copper
foils installed for {\it in-situ} X-ray energy calibration.

There are three background X-ray lines overlapping with the
kaonic-hydrogen signals : kaonic oxygen 7-6 (6.0 keV), kaonic nitrogen
6-5 (7.6 keV) and the characteristic X ray of copper $K\alpha$ (8.0 keV).
In the fitting procedure of the kaonic-hydrogen spectrum,
it turned out to be essential to use the kaonic-deuterium spectrum 
to quantify the kaonic background X-ray lines.
Therefore, we performed a simultaneous global fit
of the hydrogen and deuterium spectra,
where the intensities of the background X-ray lines
were determined using both spectra
and a normalization factor defined by the intensity ratio
of the high-statistics kaonic-carbon 5-4 peak seen in both spectra.
In Fig.~\ref{fig:SpecFit} (b) and (c),
the resulting fit lines are shown
together with components of both the background X-ray lines
and a continuous background;
(a) shows the residuals of the measured kaonic-hydrogen X-ray spectrum
after subtraction of the fitted background,
clearly displaying the kaonic-hydrogen $K$-series transitions.

As a result, the $1s$-level shift $\epsilon_{1s}$ and width
$\Gamma_{1s}$ of kaonic hydrogen were determined by SIDDHARTA to be
\begin{eqnarray*}
\epsilon_{1s} = -283 \pm 36 \rm{(stat)} \pm 6 \rm{(syst) ~eV} \\
\rm{and}~~ \Gamma_{1s} = 541 \pm 89 \rm{(stat)} \pm 22 \rm{(syst) ~eV},
\end{eqnarray*}
respectively,
where the first error is statistical and the second is systematic.
The quoted systematic error is a quadratic summation
of the following contributions :
the SDD gain shift,
the SDD response function,
the ADC linearity,
the low-energy tail of the kaonic-hydrogen higher transitions,
the energy resolution,
and the procedural dependence shown by an independent analysis
\cite{SIDT_KH}.

\section{Conclusion}
\label{sec:concl}

\begin{figure}[t]
 \begin{center}
  \includegraphics*[width=1.0\linewidth]{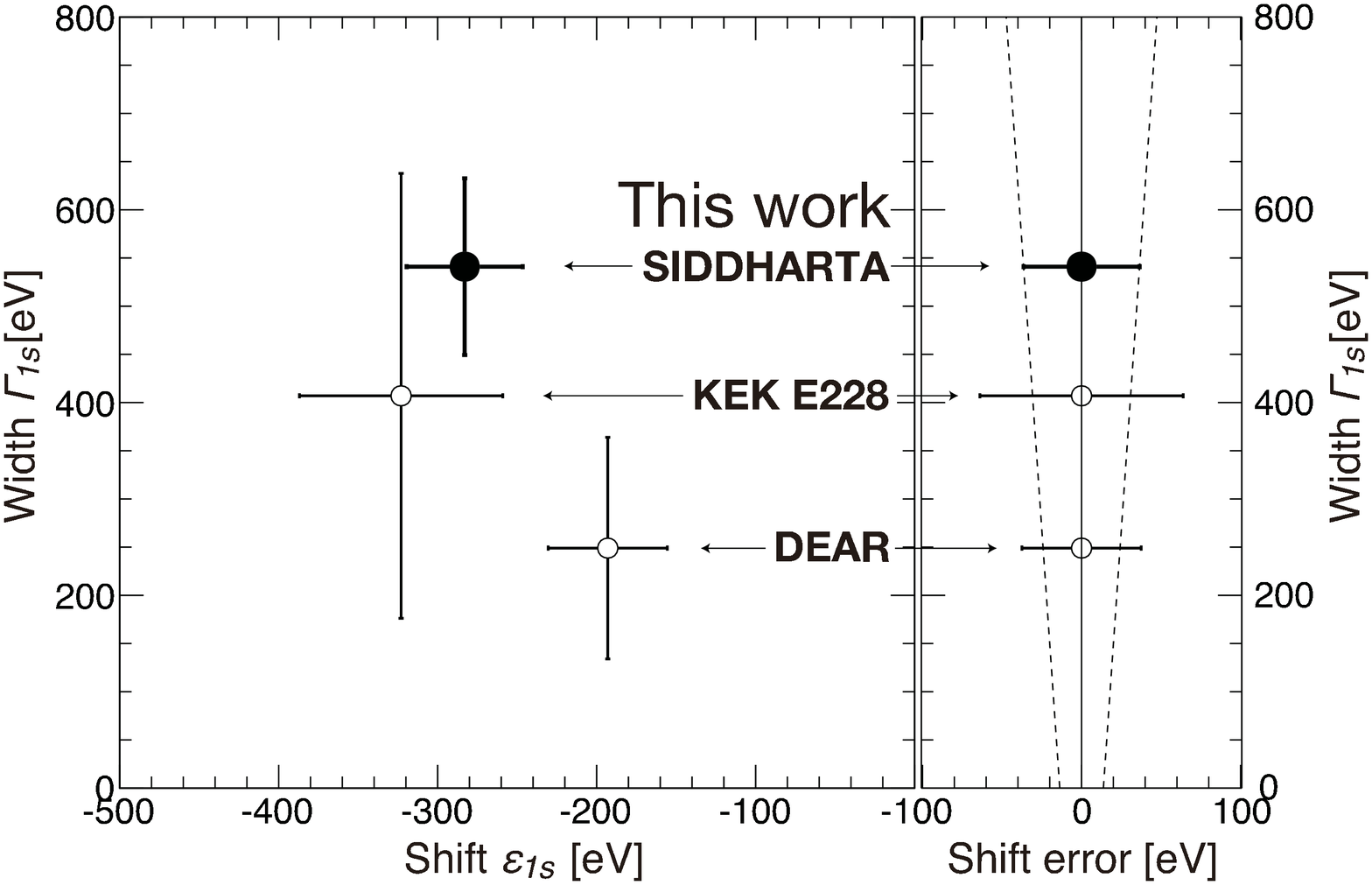}
  \caption{Comparison of the present result for the strong-interaction
  $1s$-energy-level shift and width of kaonic hydrogen
  with the two experimental results :
  KEK-PS E228 (1997) \cite{kpx} and DEAR (2005) \cite{dear}.
  The error bars correspond to quadratically added statistical and systematic
  errors.
  The right panel shows
  the error in the energy shift as a function of the width (vertical axis)
  for each experiment.
  The dashed lines represent the SIDDHARTA precision
  calculated assuming the same statistics but with differing width.
  }
  \label{fig:ShiftErrWidth}
 \end{center}
\end{figure}

We have determined the strong-interaction energy-level
shift and width of the kaonic-hydrogen atom $1s$ state
with the best accuracy up to now \cite{SIDT_KH}.
The obtained shift and width are plotted in Fig.~\ref{fig:ShiftErrWidth}
along with the other two recent results \cite{kpx, dear}.
It should be noted that the smaller the width,
the better the accuracy of determining the energy.
The right panel of Fig.~\ref{fig:ShiftErrWidth} shows
the errors on the energy shift as a function of the width (vertical axis)
for each experiment,
together with guide lines representing SIDDHARTA precision
calculated assuming the same statistics but with differing width.
In comparison with the DEAR result,
the accuracy of determining the energy in SIDDHARTA is
obviously improved.

The new triggerable X-ray detectors, SDDs,
developed in the framework of the SIDDHARTA project,
lead to an improved energy and time resolution over the past
experiments,
resulting in much lower background in comparison with the DEAR
experiment, thus permitting a better control on systematics.

Our determination of the shift and width allows
more precise evaluation of $\overline{K}N$ scattering lengths
which yields vital constraints on the theoretical description
of the low-energy $\overline{K}N$ interaction 
($e.g.,$ \cite{Cieply2011,Ikeda_2011}).

For further study of the ``isospin-dependent'' $\overline{K}N$ interaction,
it is crucial to measure the strong-interaction $1s$-energy-level
shift and width of kaonic deuterium.
The present kaonic-hydrogen result combined with kaonic-deuterium data to be
collected in the SIDDHARTA-2 experiment \cite{SIDT2},
being designed with more than one order-of-magnitude
better signal-to-background ratio than that in SIDDHARTA,
will provide invaluable constraints for the theories
of low-energy QCD in the strangeness sector.

\section*{Acknowledgments}
\label{sec:ack}

We thank C. Capoccia, B. Dulach, and D. Tagnani from LNF-INFN; and
H. Schneider, L. Stohwasser, and D. St$\ddot{\mbox{u}}$ckler
from Stefan-Meyer-Institut,
for their fundamental contribution in designing and building the
SIDDHARTA setup.
We thank as well the DA$\Phi$NE staff for the excellent working
conditions and permanent support.
Part of this work was supported by
HadronPhysics I3 FP6 European Community program,
Contract No. RII3-CT-2004-506078;
the European Community-Research Infrastructure Integrating
Activity ``Study of Strongly Interacting Matter''
(HadronPhysics2, Grant Agreement No. 227431)
under the Seventh Framework Programme of EU;
Austrian Federal Ministry of Science
and Research BMBWK 650962/0001 VI/2/2009;
Romanian National Authority for Scientific Research (ANCS);
and the Grant-in-Aid for Specially Promoted Research (20002003), MEXT, Japan.





\bibliographystyle{elsarticle-num}
\bibliography{<your-bib-database>}



\end{document}